# Diode Like Attributes in Magnetic Domain Wall Devices via Geometrical Pinning for Neuromorphic Computing

Rahaman Hasibur[1], Durgesh Kumar[1], Chung Hong Jing[2], Maddu Ramu[1], Lim Sze Ter[2], Tianli Jin[1], S. N. Piramanayagam[1, *]

[1]School of Physical and Mathematical Sciences, Nanyang Technological University, 21 Nanyang Link, Singapore, 637371

[2]Institute of Materials Research and Engineering, A*STAR, 2 Fusionopolis Way, Innovis, Singapore, 138634

[*]*prem@ntu.edu.sg*

## Abstract

Neuromorphic computing (NC) is considered as a potential vehicle for implementing energy-efficient artificial intelligence (AI). To realize NC, several materials systems are being investigated. Among them, the spin-orbit torque (SOT) -driven domain wall (DW) devices are one of the potential candidates. To implement these devices as neurons and synapses, the building blocks of NC, researchers have proposed different device designs. However, the experimental realization of DW device-based NC is only at the primeval stage. In this study, we have proposed and investigated pine-tree-shaped DW devices, based on the Laplace force on the elastic DWs, for achieving the synaptic functionalities. We have successfully observed multiple magnetization states when the DW was driven by the SOT current. The key observation is the asymmetric pinning strength of the device when DW moves in two opposite directions (defined as, $x_{hard}$ and $x_{easy}$). This shows the potential of these DW devices as DW diodes. We have used micromagnetic simulations to understand the experimental findings and to estimate the Laplace pressure for various design parameters. The study leads to the path of device fabrication, where synaptic properties are achieved with asymmetric pinning potential.

## Introduction

Recently, AI has gained popularity in a variety of areas ranging from consumer devices such as smartphones and televisions to manufacturing industries [1-4]. Despite the great demand, AI needs better solutions to the enormous power consumption associated with its implementation [1-3]. At present, AI is being realized in existing computing devices based on the von Neumann architecture [2, 3, 5]. This architecture suffers from the problems of operational speed and power consumption [5]. To overcome these limitations, researchers took inspiration from the highly energy-efficient and compact "human brain" and attempted to emulate its neurobiological features. Recently, brain-inspired computing, or neuromorphic computing (NC) architecture has sparked tremendous interest due to its potential in low-energy computation [1-3, 5]. Like the human brain, NC architecture too consists of many neurons and synapses. Neurons act as processors, while synapses store the information. In addition, synapses act as a bridge between the pre-neuron and the post neurons and controls the weight of the travelling information. Therefore, synapses must exhibit multi-conductance (resistance) states in synthetic synaptic devices [2, 3].

Different material classes have been investigated to realize artificial synapses and neurons. Out of them, resistive random-access memory [6], ferroelectric materials [7], phase-change materials [8], and spintronics materials [1-3, 5] have gotten much more attention in the research community. Recent research indicates that spintronic domain wall (DW) devices are potentially useful for NC owing to their low energy consumption, high endurance, and non-volatile nature [2, 9]. In 2012, Sharad et al. simulated the DW-based synapse in their spin-based synapse-neuron model intended for character recognition tasks [10]. Initial DW motion studies were based on the spin-transfer-torque (STT) effect, which gets generated when the electrical current flows through the ferromagnetic layer [11-13]. Around a decade ago, it was found that the current density for driving the DWs or switching the magnetization can be reduced at least by a factor of 10 by utilizing the spin-orbit torque (SOT) from the heavy metal layer, deposited adjacent to the ferromagnetic layer [14-18]. Therefore, Sengupta et al. in their simulation work emulated the functionalities of energy-efficient neurons and synapses using SOT-driven DW-MTJ device [19-21].

For the practical realization of DW devices, it is essential to understand the nature of DW motion. As DW motion is of a stochastic nature, controlling the DW motion is of the utmost importance. To achieve the controlled, reliable, and reproducible DW motion researchers studied different mechanisms to pin the DWs at desired sites [2, 22]. These are termed artificial pinning sites and can be of geometrical or non-geometrical (or magnetic) origin [2, 3, 22-29]. Several such pinning sites could be used to achieve multiple magnetization (resistance) states in DW-MTJ devices. In earlier days of DW device research, the artificial pinning sites in the shape of triangular notches were introduced to demonstrate the DW pinning [30]. For instance, Kläui et al. showed the DW pinning using triangular notches in the ring-shaped NiFe wires by utilizing Giant-magneto-resistance (GMR) measurement [31]. As an extension, Parkin et al. fabricated multiple triangular notches along straight NiFe wire and demonstrated the successive DW pinning at all the pinning sites by using magnetic force microscopy (MFM) [32]. Later the concept of triangular, rectangular, and circular anti-notches was also explored. Besides, the concept of the stepped nanowire was studied to pin the DWs [28]. Kumar et al. have also explored DW pinning in meander devices where DWs were driven by SOT in W/CoFeB/MgO stack [9]. They also demonstrated the multi-resistance states in these devices using anomalous Hall resistance [33].

In addition to geometrical pinning mechanisms, a few non-geometrical methods were also studied. For instance, Polenciuc et al. utilized the concept of exchange bias for creating the pinning sites [34]. Subsequently, Jin et al. studied the concepts of non-magnetic metal diffusion, ion-implantation of non-magnetic $B^+$ ions, and the exchange coupling between in-plane and out-of-plane magnetized ferromagnetic wires to fabricate the synthetic pinning sites [31, 35-41]. Despite the above successes in DW device research, the experimental realization of DW-device-based neuromorphic computing is only at the primordial stage. Therefore, more methods to fabricate synthetic neurons and synaptic elements need to be researched. In this paper, we have studied the concept of pine-tree-shaped DW devices for achieving multiple magnetization states and hence, multiple-resistance states.

The pine-tree DW synaptic device is the repetitions of a trapezium with a constant gradient in the width of the segment. Figure 1 (a) shows the schematic diagram of the proposed DW device. When the DW reaches the constriction, it requires a higher energy to depin from the pinning site, due to the Laplace pressure (and hence, Laplace force) on the elastic DW [42]. As the proposed pinning site structure is asymmetric in shape, we anticipate that the Laplace force should also be antisymmetric. As a result, asymmetric DW motion may be expected when driven along $x_{easy}$ or $x_{hard}$ directions. Here, $x_{easy}$ and $x_{hard}$ are defined along +x and -x directions, respectively. This way, one may expect the multiple resistance states along both the above-mentioned directions. However, the nature of the DW surface is expected to differ slightly, resulting in a diode-like nature of the DW motion which can also be useful in spin-based logic applications. This pine tree device has also the potential to mimic "in synapse neuron function" due to its unique energy landscape.

## Micromagnetic Simulations

To understand the proposed concept, we first performed micromagnetic simulations using Mumax[3] [43]. The simulations utilize the Landau Lifshitz Gilbert (LLG) equation [43, 44]. The geometric and magnetic parameters used during the simulations are listed in Table 1. These magnetic parameters can experimentally be realized in W/ CoFeB/ MgO system [12, 45, 46]. In simulations, we used SOT to drive the DWs. Moreover, the *i*DMI breaks the required symmetry for the field-free DW motion. In all the simulations, we first introduced a DW at one end of the wire, and then pushed further along the DW device using DC/ pulsed current. In our simulations, blue and red colors represent the magnetizations along -z and +z directions, respectively.

We first simulated the DW motion in the reference (straight nanowire) DW device. Figure 1 (b) shows the DW motion along $x_{easy}$ and $x_{hard}$ under the influence of SOT current. We investigated DW dynamics for a large range of applied current densities ranging from $1\times10^{10}$ A/m$^2$ to $9\times10^{11}$ A/m$^2$. However, we have only shown a few selected results in figure 1 (b) for clarity. The $m_z$ vs $t$ plots for all the other current densities are presented in supplementary section 1. In all the cases, the DW moves from one end of the nanowire to the other without any DW pinning. Here, we have utilized the *z*-component of magnetization ($m_z$) to represent the DW position. In addition, we have observed an increment in the slopes of $m_z$ vs $t$ plots with the current density. This suggests that the DW velocity increases as current density increases and gets saturated beyond a *J* of $2\times10^{10}$ A/m$^2$. As expected, all these results are identical for both $x_{easy}$ and $x_{hard}$ directions. This illustrates the symmetry in the DW dynamics in the reference wire.

***Table 1:*** *Geometric and Magnetic parameters used during micromagnetic simulation. Here, $l_s$, w, and t represent the length, width, and thickness parameters, respectively.*

| Parameter | Value |
| --- | --- |
| Dimensions (reference device) "$l \times w \times t$" | 1500 nm×50 nm×1 nm |
| Dimensions (proposed device) "$w_1$, $w_2$, $l_s$, $t$, no. of segments, gradient of segment, $w_1$: $w_2$" | 75 nm, 50 nm, 75 nm, 1 nm, 10, 0.17, 1.5:1 |
| | 100 nm, 50 nm, 100 nm, 1 nm, 10, 0.25, 2:1 |
| | 150 nm, 50 nm, 150 nm, 1 nm, 10, 0.33, 3:1 |
| Cell size | 1 nm×1 nm×1 nm |
| Exchange length ($l_{ex}$) | 4.89 nm |
| Spin Hall angle ($\theta_{SH}$) | 0.5 |
| Damping constant ($\alpha$) | 0.012 |
| Exchange constant ($A_{ex}$) | $1.5 \times 10^{-11}$ J/m |
| Saturation magnetization ($M_s$) | $1 \times 10^6$ A/m |
| Interfacial Dzyaloshinskii Moriya interaction ($i$DMI) constant ($D$) | 0.5 mJ/m$^2$ |
| Anisotropy constant ($K_u$) | $1 \times 10^6$ J/m$^3$ |
| Anisotropy direction | (0, 0, 1) |

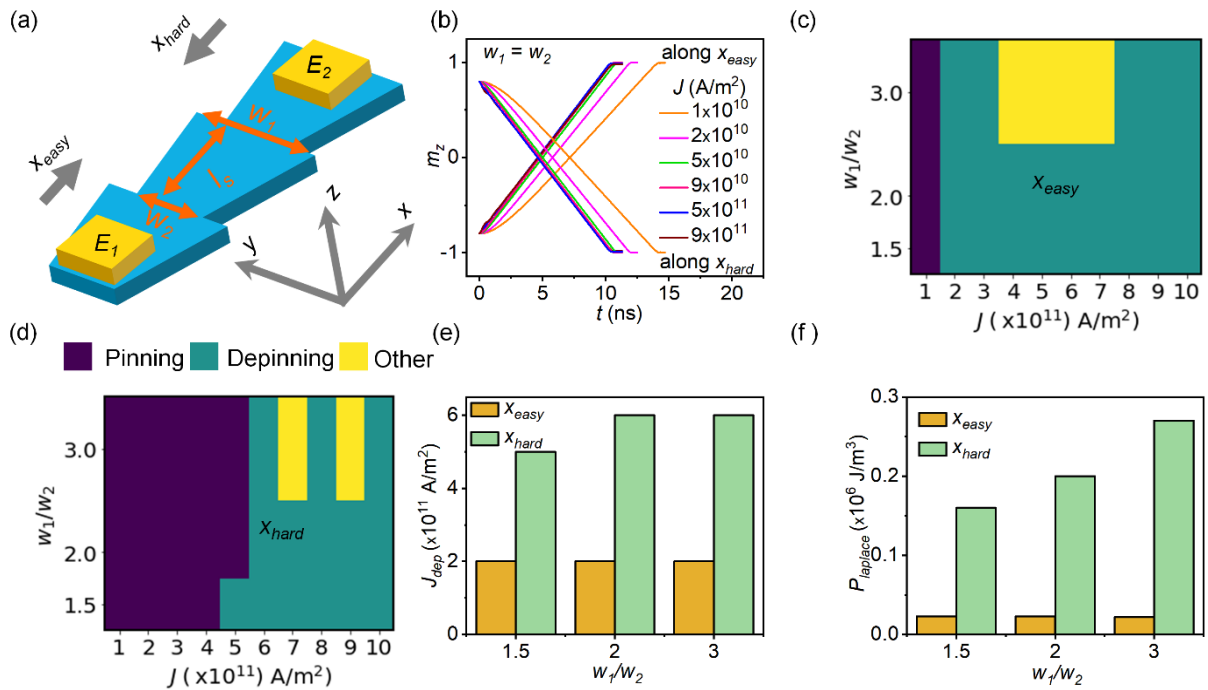

*Figure 1. (a) The schematic illustration of pine-tree DW-based synaptic devices. The segment's wider end, narrower end, and segment length are represented by $w_1$, $w_2$, and $l_s$, respectively. The directions along +x and -x are referred to as $x_{easy}$ and $x_{hard}$, respectively. (b) The plot of domain wall position ($m_z$) vs simulation time (t) for DW motion along $x_{easy}$ as well as $x_{hard}$ directions. Current densities in a range from $1 \times 10^{10}$ A/m$^2$ to $1 \times 10^{12}$ A/m$^2$ were applied. The phase diagram of DW dynamics as a function of J and $w_1/w_2$ along (c) $x_{easy}$ and (d) $x_{hard}$. The variation of (e) depinning current density ($J_{dep}$) and (f) Laplace pressure ($P_{laplace}$) as a function of $w_1/w_2$ for DW motion in both directions.*

As a next step, we studied SOT-driven DW dynamics in the proposed pine tree synaptic devices. A similar range of current densities was used in these simulations as well. The observed DW dynamics can be broadly divided into three regions as a function of $J$ and $w_1/w_2$. The regions are referred to as (*i*) pinning region (*ii*) depinning region, and (*iii*) other. For DW motion along $x_{easy}$, the DW gets pinned at the first pinning for all the $J$ values smaller than $2\times10^{11}$ A/m$^2$ at all values of $w_1/w_2$ (1.5, 2, and 3). This is called the pinning region and is represented by the violet color. Above a current density of $2\times10^{11}$ A/m$^2$, the DW gets depinned from the pinning site for all $w_1/w_2$ values. This is defined as the depinning region and is represented in olive color. Besides these two, we also observed the random nucleation of reversed domains (predominantly near the pinning sites) for large values of $J$ and $w_1/w_2$. These are included in 'other' category and is shown in yellow color. We have presented the details of these random nucleation in supplementary section 2.

Similarly, we performed the simulations for the DW motion along the $x_{hard}$, and the phase diagram of DW motion is presented in figure 1 (d). Similar to the case of DW motion along $x_{easy}$ direction, the present case also exhibits the three regions in the phase diagram. The pinning region expands for this case indicating stronger pinning in $x_{hard}$ direction. We found similar anomalies in DW dynamics in terms of random nucleation at higher current density values for devices with $w_1/w_2 = 3$. We can also observe that the pinning-depinning boundary is not a straight line. A careful inspection of the phase diagram leads to the conclusion that the depinning current density becomes independent of $w_1/w_2$ for motion in $x_{easy}$ direction. However, we found a change in depinning current density with the values of $w_1/w_2$ when the motion is along $x_{hard}$. The above observations are demonstrated in figure 1 (e) as a variation of depinning current density ($J_{dep}$) with $w_1/w_2$. These results indicate the asymmetry in the pinning strength of the pinning sites along the two above-mentioned directions.

To get a deeper insight into its physics, we performed the simulations for a single pinning site. Firstly, we determined the $J_{dep}$ for the pine-tree DW devices consisting of a single pinning site with different $w_1/w_2$ values. For the DW motion along both $x_{easy}$ and $x_{hard}$, we found exactly a similar trend of $J_{dep}$ as a function of $w_1/w_2$ as observed in figure 1 (e). These results are presented in supplementary section 3. This $J_{dep}$ is the current density at which the external torque due to SOT current overcomes the torque due to Laplace force concerning the single pinning site. We have calculated Laplace pressure for both $x_{easy}$ and $x_{hard}$ directions. The Laplace pressure acting on the DW is inversely proportional to the radius of curvature according to the formula $P_{laplace} = \frac{\gamma}{r_c}$ [42, 47-49]. Here, $\gamma$ is the DW surface energy density and $r_c$ is the radius of curvature of the DW surface at the pinning sites. We found that $r_c$ in the case of $x_{hard}$ is far smaller than that observed for $x_{easy}$. The details of the $P_{laplace}$ estimation are included in the supplementary section 4. Figure 1 (f) depicts the variation of $P_{laplace}$ with $w_1/w_2$ for both $x_{easy}$ as well as $x_{hard}$ directions. Along $x_{easy}$, $P_{laplace}$ becomes independent of $w_1/w_2$, which is consistent with the fact that the $J_{dep}$ is uniform for all the $w_1/w_2$ cases. On contrary, for the DW motion along $x_{hard}$, $P_{laplace}$ has an increasing trend. For all the cases, $P_{laplace}$ is larger along $x_{hard}$ than that along $x_{easy}$. This again confirms the asymmetric nature of proposed pine-tree synaptic devices.

Subsequently, we studied the successive pinning and depinning at all the pinning sites to achieve intermediate magnetization states (multi-resistance states) in the pine-tree synaptic devices. We applied pulsed SOT current in such a way that one pulse pushes the DW from one pinning site to the next one. Once the DW reaches a pinning site, we switched off the current and observed the nature of DW dynamics. For $w_1/w_2 = 1.5$, we applied the pulses of a current density of $5\times10^{11}$ A/m$^2$. More details on the parameters of pulsed current density are presented in supplementary section 5. In response to such a pulsed current density, we achieved the multiple resistance states for the DW motion along both the directions (figure 2 (a)). Usually, the magnitude of the current density in these experiments is equivalent to or higher than the depinning current density. Similar results were obtained for $w_1/w_2$ = 2 and 3 (figures 2 (b & c)). In both these cases, a current density of $6\times10^{11}$ A/m$^2$ was used. It is notable that a staircase-like structure could not be observed for the DW motion along $x_{hard}$ for the devices with $w_1/w_2$ = 3. This is because very high pinning strength results in the random nucleation of reversed domains at the pinning sites. Therefore, a proper DW motion could not be observed. We observed DW fluctuations in constant $m_z$ states (when $J = 0$) for the devices with the lowest $w_1/w_2$ value. However, this fluctuation diminishes, and DW is firmly pinned for devices with higher $w_1/w_2$. This feature is related to the strength of the pinning potential at the pinning sites. To have a visual understanding of the intermediate magnetization states, we have presented the snapshots of instantaneous magnetization in different pinning events. Figures 2 (d & e) represent the corresponding results for $x_{easy}$ and $x_{hard}$ directions, respectively. When DW reaches the pinning site, it expands (following the sudden change in the width) under the influence of the SOT current. Simultaneously, the surface tension of the magnetic bubble tries to restore the shape of the bubble [42, 47]. As a result, there is a Laplace force in a direction opposite to the DW motion [42, 47]. This way the DW pinning is achieved.

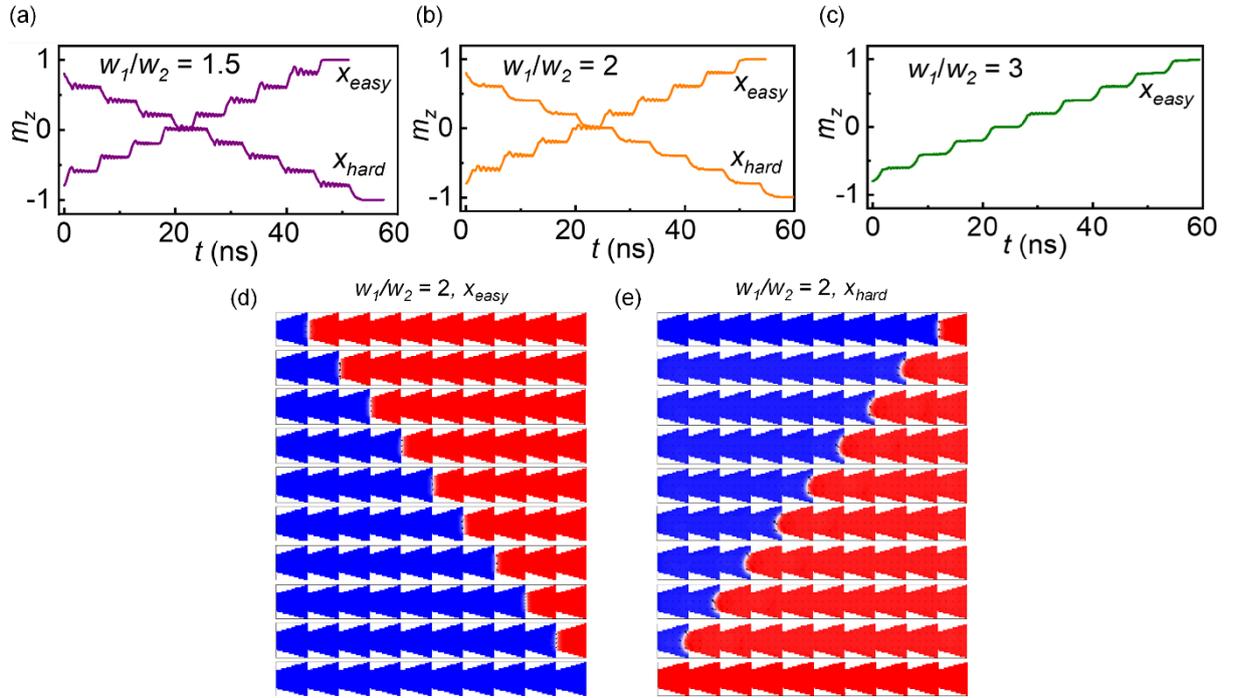

*Figure 2.* The illustration of multiple-resistance states for DW motion in both the directions for $w_1/w_2$ of (a) 1.5, (b) 2, and (c) 3. The snapshots of DW pinning in devices with $w_1/w_2$ = 2 when DW moves along (d) $x_{easy}$ (e) $x_{hard}$.

## Experimental Results
## Thin Film Deposition

The sample stack Si/SiO$_2$ substrate/ HP-W (3 nm)/ LP-W (3 nm)/ Co$_{40}$Fe$_{40}$B$_{20}$ (1 nm)/ MgO (1 nm)/ Ru (2 nm) was deposited using the DC/RF sputtering. Please refer to the inset of figure 3 (a) for the schematic of the studied thin film stack. To confirm perpendicular magnetic anisotropy (PMA) in our samples, we measured the magnetic hysteresis (M-H) loops using polar Kerr microscopy, as shown in figure 3 (a). The M-H loop revealed PMA in our samples at an as-deposited state and the coercivity of our samples was found to be 45 Oe. We have also measured the SOT properties of these samples and observed reasonably good SOT characteristics [9].

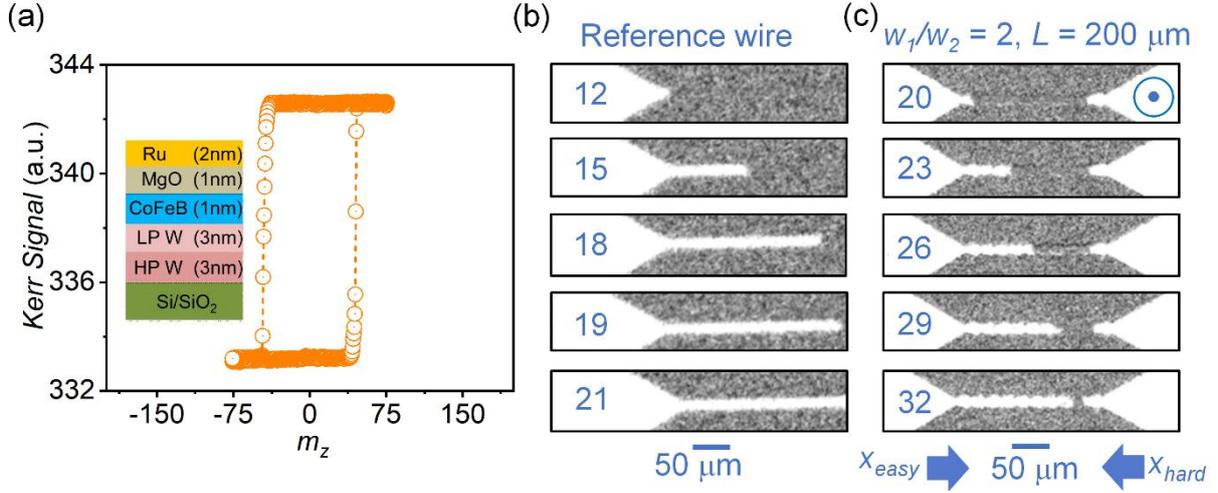

*Figure 3.* (a) The magnetic hysteresis (M-H) loop captured using polar Kerr microscopy, in the out-of-plane (OOP) direction. The M-H loop shows perpendicular magnetic anisotropy (PMA) in the studied dual W sample. Inset: The thin film stack used in this paper. OOP magnetic field ($H_z$)-driven domain wall motion in (b) the straight (or reference) wire of width 10μm and length 200 μm and (c) pine tree devices. The digits in blue on the top of Kerr images represent the number of field pulses (e.g., $n^{th}$ field pulse) applied before capturing the corresponding magnetization state.

## Nature of the Domain Wall Motion

In the first step, we studied the DW motion as a function of the out-of-plane (OOP) magnetic field in the reference wires. We first saturated the devices with a large OOP magnetic field of -500 Oe. the white Kerr signal represents the magnetization along the +z-direction. Subsequently, we applied a reversed OOP magnetic field of 50 Oe in the form of pulses. A pulse width of 1 s was utilized. As one can see in figure 3 (b), after the application of the 12$^{th}$ field pulse, the reversed domain gets nucleated and the DW reaches the neck of the nucleation pad and microwire. The DW moves from the left end of the wire to the right under the influence of the subsequent field pulses. As expected, no extrinsic DW pinning was observed. Later, we repeated these experiments on the pine tree synaptic devices. These results are presented in figure 3 (c). The DW motion was observed along the $x_{easy}$ direction. However, DW does not move along $x_{hard}$. If we compare the results of figures 3 (b) and 3 (c), we find the following: The DW velocity ($v$) in the reference wire is 24.2 μm/s. However, the same in the case of the pine tree device is 11.7 μm/s along $x_{easy}$ and almost zero along $x_{hard}$ direction. This suggests the $x_{easy}$ direction in pine tree devices offers DW pinning but the DW pinning is smaller. However, the same along $x_{hard}$ is much larger. These results are in agreement with our results in micromagnetic simulations.

Subsequently, we performed the SOT-driven DW motion experiments in pine-tree synaptic devices for probing the DW motion. These results are presented in figure 4. We first saturated the devices using a large OOP magnetic field. After inserting the DW into the DW devices, we applied the pulsed current of various amplitudes. The pulse width and time interval between consecutive pulses were taken as 7 ms and 2.5 s. Such a large time interval was used to avoid the Joule heating in our devices. A constant longitudinal magnetic field ($H_x$) of 200 Oe was applied in a direction as shown in figure 4 (along +$x$ direction) to induce the deterministic DW motion. The direction of the $J$ was kept fixed along the -$x$ direction in all the experimental results presented in this figure. We considered maximum current density based on the cross-sectional area at the narrower region $w_2$, and we mentioned it as $J_{max}$. We optimized the $H_x$ value of 200 Oe as an optimized value to induce the SOT-induced DW motion. A larger $H_x$ results in faster DW motion (slightly difficult to control) and a smaller $H_x$ results in a much slower DW motion (larger experimental duration). All the experiments were performed for the DW motion along $x_{easy}$. We found uncontrolled DW motion along $x_{hard}$, when driven only with SOT. Later, we assisted the SOT with $H_z$ and observed controlled and repeatable DW motion. We will discuss these results in the next sub-section.

For the edge ratios ($w_1/w_2$) of 1.5, we could not observe any DW motion (or only a marginal DW motion) for the current densities of $1\times10^{10}$ A/m$^2$ (1 mA) and $2\times10^{10}$ A/m$^2$ (2 mA). This could be because of (*i*) creep DW motion at such low current densities or (*ii*) pinning offered by the pinning sites or (*iii*) both. Once the current density was increased to $3\times10^{10}$ (3 mA) and $4\times10^{10}$ A/m$^2$ (4 mA), DW motion was observed (as shown in figures 4 (a) and (b)) from one end of the devices to the other in multiple current pulses. A further increase of current density to $5\times10^{10}$ A/m$^2$ (5 mA) and higher results in sharp DW motion from one end to the other. This is also shown in the narrowest violet (pinning region) and broadest yellow (depinning region) region in the phase diagram of DW motion as a function of $w_1/w_2$ and $J_{max}$ (figure 4 (f)). We have included the current densities for which negligible DW motion was observed in the pinning region. Based on these results, we conclude that the pinning offered by the devices with $w_1/w_2$ of 1.5 is negligibly small.

Later, we performed these experiments for the devices with an edge ratio of 2. In this case, we did not observe any DW motion (or negligible DW motion) for current density values of $1\times10$ (1 mA), $2\times10^{10}$ (2 mA), and $3\times10^{10}$ A/m$^2$ (3 mA). For $J_{max} = 4\times10^{10}$ A/m$^2$ (4 mA), a relatively larger pinning strength was observed (figures 4 (c) and (d)). Similar DW motion was observed for the current densities ranging from $5\times10^{10}$ A/m$^2$ (5 mA) to $8\times10^{10}$ A/m$^2$ (8 mA). Besides DW motion, nucleation of reversed domains at the pinning sites was also observed at larger current densities. This is because of relatively larger SOT efficiency and Joule heating at the pinning sites [50, 51]. This aspect was also highlighted in our micromagnetic simulations. At current densities of $9\times10^{10}$ A/m$^2$ (9 mA) and above, a sharp DW motion was observed. As can be seen in figure 4 (f), the pinning region broadens (and the depinning region shrinks) for this case. From these results, we can say that an improved pinning strength was observed for $w_1/w_2$ of 2.

Lastly, we performed these experiments for the pine-tree devices with $w_1/w_2 = 3$. For this case, we observed the nucleation of the reversed domains for most of the studied current densities. Moreover, the pinning strength becomes too large to depin the DWs from the pinning sites. We have shown the results of $J_{max} = 1.2\times10^{11}$ A/m$^2$ (12 mA) in figure 4 (e). As a result, we could not observe the full coverage reversed magnetization all over the device.

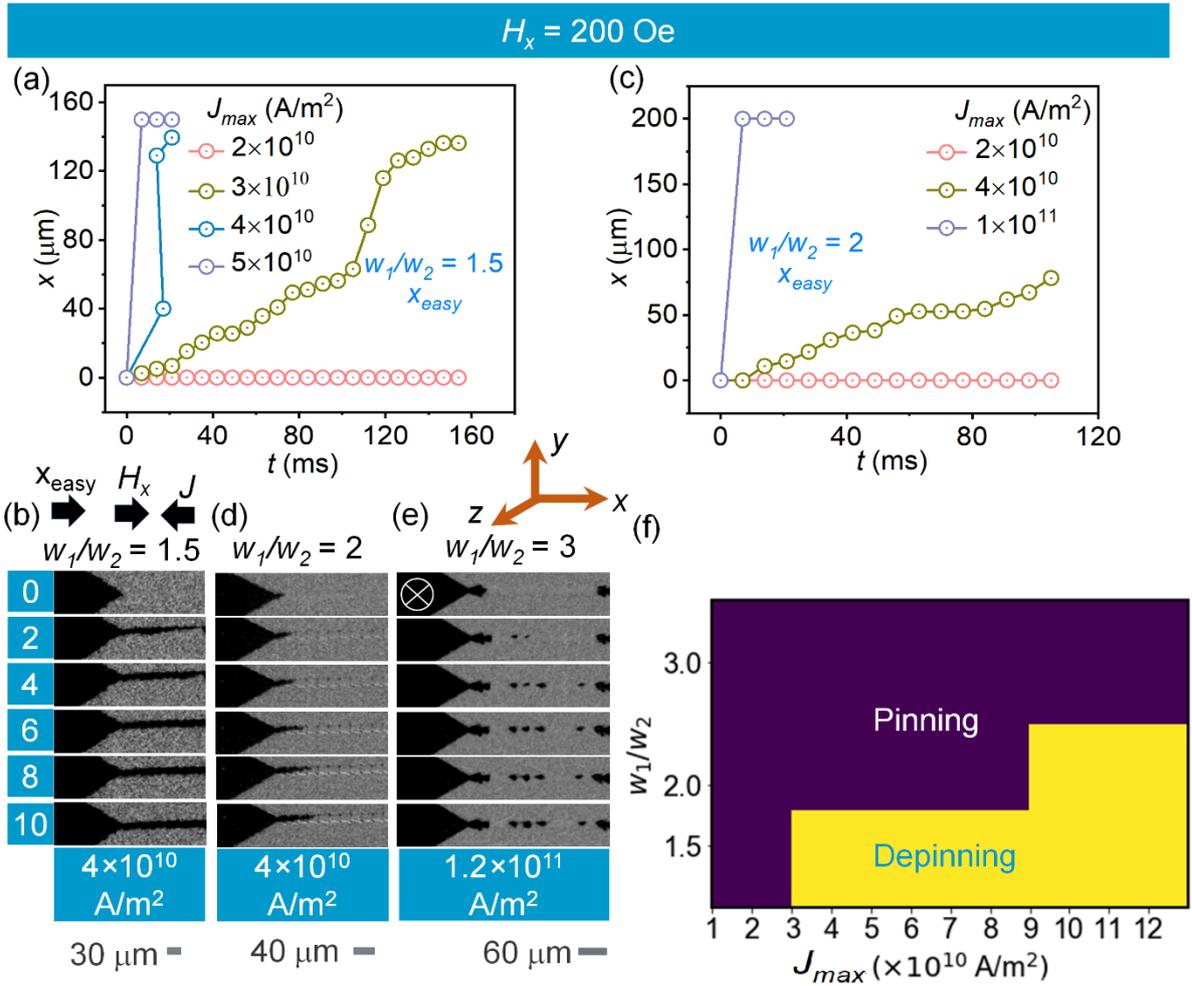

*Figure 4.* SOT-driven DW motion in pine-tree synaptic devices for edge ratios ($w_1/w_2$) of (a) 1.5 and (b) 2 for different current density values. The corresponding current density values are listed in the graphs. Kerr microscopy images showing the nature of DW motion for (c) $w_1/w_2 = 1.5$ and $J_{max} = 4 \times 10^{10}$ A/m$^2$, (d) $w_1/w_2 = 2$ and $J_{max} = 4 \times 10^{10}$ A/m$^2$, and $w_1/w_2 = 3$ and $J_{max} = 1.2 \times 10^{11}$ A/m$^2$. (f) Phase diagram of DW motion as a function of $w_1/w_2$ and $J_{max}$. The pinning region expands as the edge ratio increases. Here, violet, and yellow colors represent the pinning and depinning regions, respectively. All these experiments were performed for the DW motion along $x_{easy}$. In all the results of this panel, a longitudinal magnetic field of 200 Oe was applied. The directions of $J$ and $H_x$ are identical in all the results and are shown in the panel.

Intermediate Magnetization States: Multiple-resistance States

Based on the above results, we chose the devices with $w_1/w_2$ of 2 for further synaptic measurements. However, the nucleation of the reversed domains may not be desired for ideal DW-based synaptic devices. Therefore, to avoid this issue, we made the following modifications to the devices: (a) We reduced the pulse width of the devices to effectively reduce the Joule heating (particularly at the pinning sites). (b) We slightly increased the length of the segments (from 20 μm to 30 and 50 μm) while keeping the $w_1/w_2$ at 2. This effectively changes the rate of change of pinning strength in every segment. Moreover, this provides better experimental controllability on DW motion.

We performed experiments to study the existence of intermediate magnetization states (corresponding to multi-resistance states in a magnetic tunnel junction device) in our synaptic devices with modified dimensions. The results of the studies as a function of the OOP magnetic field are presented in figure 5 (a-b). The DW moves from the left end of the device to the right with some hint of pinning at the pinning sites. However, as we can see in figure 5 (a-b), the pinning is not very strong. Nevertheless, we observed a total of 9 distinguishable intermediate magnetization states (including the completely up and down magnetization states). These states are highlighted with the stars in figure 5 (b). Note, that these experiments were performed for the DW motion along the $x_{easy}$ direction. We also attempted the same for $x_{hard}$ direction. But the DW motion could not be controlled in this direction. These results suggest that the magnetic field may not be a suitable stimulus for realizing multiple resistance states in our devices. Therefore, we performed similar experiments with SOT.

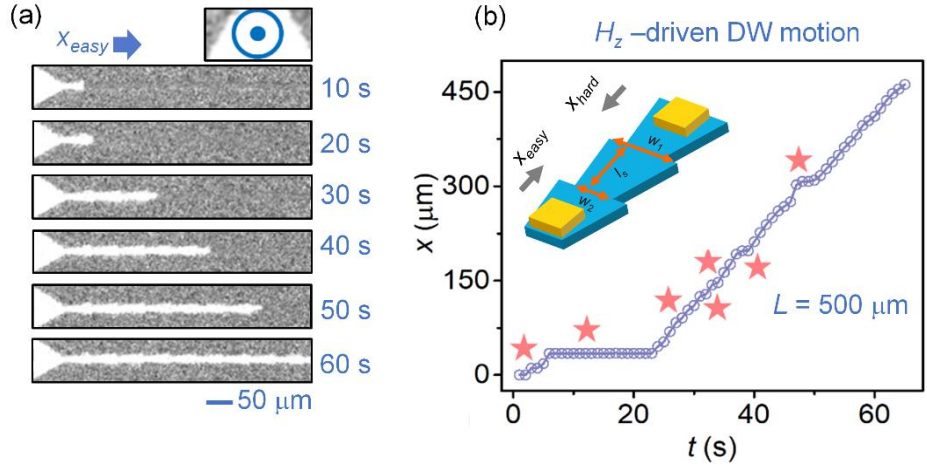

*Figure 5.* (a) The Kerr microscopy images of the magnetization state of our synaptic devices when DW is moved along the $x_{easy}$ direction. The number at the right of these images is the experimental time. Here, the white color represents the magnetization along the +ve z-axis. (b) The DW position (x) vs time (t) graph that shows nine different magnetization states. L represents the total length of the above pine tree device.

For deriving intermediate magnetization states when the DW is driven by SOT, we applied the pulsed current of $1\times10^{11}$ A/m$^2$. A longitudinal magnetic field of 400 Oe was utilized. The directions of J and $H_x$ are indicated in figure 6. For a magnetic field ($H_x$) larger than this random nucleation of the oppositely magnetized domains starts appearing at the pinning sites. While for magnetic fields smaller than this value, DW becomes very slow, resulting in a very large experiment time. For the DW motion along $x_{easy}$, the current density, and longitudinal magnetic field both were applied along the -x direction. A pulse width and time interval between consecutive pulses were set at 15 ms and 2 s were utilized. Such a high time interval was utilized to avoid the Joule heating in our device. Similar to figure 5 (a), the white color represents the magnetization in the +z direction. In response to such current pulses, a total of nine magnetization states were observed. For the DW motion along $x_{hard}$, we again observed 9 magnetization states. The experimental procedure employed for this case is identical except for an additional OOP magnetic field (as mentioned before). Moreover, the directions of both the current density as well as longitudinal magnetic fields were reversed. The nature of the DW motion is consistent with the SOT-driven DW motion for similar systems [9, 52]. These results are presented in figure 6 (a-b). Figure 6 (c) represents the Kerr microscopy images corresponding to the results presented in figure 6 (b). Despite the asymmetry in the DW motion along $x_{easy}$ and $x_{hard}$ directions, controlled and repeatable pinning was observed for both directions.

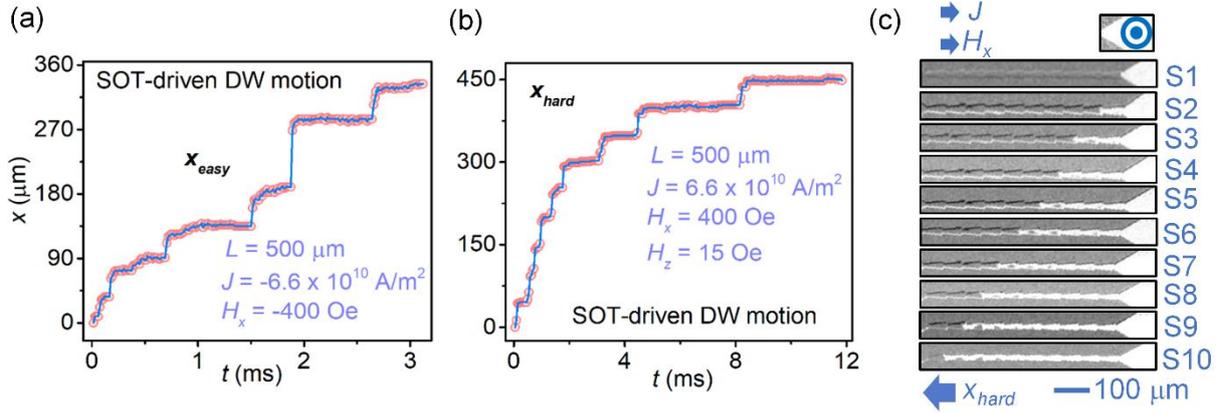

*Figure 6. The illustration of intermediate magnetization states (corresponding to multi-resistance states) in pine three device for SOT driven DW motion along (a) $x_{easy}$ and (b) $x_{hard}$ direction. (c) the Kerr microscopy images corresponding to DW motion along $x_{hard}$. Here, the white color represents the magnetization along the +ve z-axis. "Si" represents the i$^{th}$ state.*

## Conclusions

In this paper, we have demonstrated a pine-tree DW device as a synaptic element. In micromagnetic simulations, we studied DW dynamics for a vast range of current densities from $J = 1 \times 10^{11}$ to $1 \times 10^{12}$ A/m$^2$ for the DW devices having edge ratios ($w_1/w_2$) of 1.5, 2, and 3. The phase diagram of $J$ -$w_1/w_2$ clearly exhibits three regions, namely, pinning, depinning, and a nucleation (indicated as other) region. This also suggests the DW pinning for a large range of parameters. Moreover, we observed that the nature of DW motion is asymmetric for the two distinct directions of DW motion, $x_{easy}$, and $x_{hard}$. Along $x_{hard}$, DW experiences a larger pinning strength from the pinning sites compared to that along $x_{easy}$. This is because of the difference in the Laplace pressure (and hence, Laplace force) on the DW surface along the two above-mentioned directions. The Laplace pressure is largest for the device with $w_1/w_2 = 3$ and it has a detrimental effect as it corresponds to the nucleation of reversed domains at the pinning sites. Thereafter, we fabricated the micro-meter sized DW devices with different edge ratios and studied the DW dynamics under the influence of the OOP magnetic field and SOT. The field driven DW motion experiments confirm the pinning of the DWs in our devices. They also confirm the asymmetric nature of the DW motion in the two directions. A full phase diagram of DW motion as a function of $J$ and $w_1/w_2$ was studied using SOT and the DW pinning for a large range of parameters was observed. Moreover, the devices with $w_1/w_2 = 2$ exhibit the optimum pinning strength, and hence they were chosen for further study. In addition, we adjusted the geometric parameters of the devices to avoid the nucleation of the reversed domains at the pinning sites. The nature of DW motion is consistent with SOT-driven DW motion in similar reports. Moreover, our simulation and experimental results exhibit phenomenal consistency. Finally, we demonstrated 9 magnetization states in our pine tree devices, when the DW was pushed using OOP magnetic field as well as SOT. Our results contribute significantly toward the development of DW device-based NC architecture.

## Acknowledgments

The authors gratefully acknowledge the National Research Foundation (NRF), Singapore for the CRP21 grant (NRF-CRP21-2018- 0003) grant. The authors also acknowledge the support provided by Agency for Science, Technology and Research, A*STAR RIE2020 AME Grant No. A18A6b0057 for this work. RH thanks the NTU research scholarship for carrying out research at NTU.

## Conflict of Interest

The authors declare no conflict of interest.